\pdfoutput=1
\documentclass[11pt,twoside,a4paper,cmspaper,final,collab]{cms-tdr}
\def\svnVersion{f0092cb}\def\svnDate{2026/09/28}\def\cmsCernNoTag{CERN-EP-2026-103}\def\cmsCernDate{\today}\def\cmsMessage{Published in Physical Review Letters as \href{http://dx.doi.org/10.1103/g4ym-rzkt}{\doi{10.1103/g4ym-rzkt}.}}
\begin{document}\cmsNoteHeader{SUS-24-012}

\newcommand{\znunu}{\ensuremath{\PZ\to\PGn\PGn}\xspace}
\newcommand{\zmumu}{\ensuremath{\PZ\to\MM}\xspace}
\newcommand{\wlnu}{\ensuremath{\PW\to\Pell\PGn}\xspace}

\newcommand{\dmz}{\ensuremath{\Delta m^0}\xspace}
\newcommand{\dmpm}{\ensuremath{\Delta m^{\pm}}\xspace}
\newcommand{\dmpmin}{\ensuremath{\dmpm_{\text{in}}}\xspace}
\newcommand{\psig}[1]{\ensuremath{\widetilde{P}(\text{Signal}\,\vert\,\dmpmin=#1\GeV)}\xspace}
\newcommand{\pmaxsig}[1]{\ensuremath{\widetilde{P}_{\text{max}}(\text{Signal}\,\vert\,\dmpmin=#1\GeV)}\xspace}
\newcommand{\pmaxsigsimp}[1]{\ensuremath{\widetilde{P}_{\text{max}}(\text{Signal}\,\vert\,#1)}\xspace}

\newcommand{\wjets}{\ensuremath{{\PW}\mkern-2mu\text{+jets}}\xspace}
\newcommand{\zjets}{\ensuremath{{\PZ}\mkern-1mu\text{+jets}}\xspace}

\newcommand{\FastSim}{\ensuremath{\textsc{FastSim}}\xspace}
\newcommand{\FullSim}{\ensuremath{\textsc{FullSim}}\xspace}

\cmsNoteHeader{SUS-24-012}
\title{Search for electroweakinos in compressed-spectrum scenarios with low-momentum isolated tracks in proton-proton collisions at \texorpdfstring{$\sqrt{s} = 13\TeV$}{sqrt(s) = 13 TeV}}

\date{\today}

\abstract{A search for supersymmetric electroweakinos is performed using events with a low-momentum (soft) isolated track and large missing transverse momentum, targeting nearly mass-degenerate higgsino-like charginos and neutralinos. For mass splittings of 0.3--1\GeV, the chargino decays to the lightest neutralino and a low-momentum pion, which can produce a soft, potentially displaced track. A parameterized neural network separates signal from background using kinematic and impact parameter information. The analysis uses 138\fbinv of proton--proton collision data at a center-of-mass energy of 13\TeV recorded with the CMS detector. No significant excess above the standard model expectation is observed. At 95\% confidence level, the considered higgsino model is excluded for mass splittings in the range 0.28--1.15\GeV and for chargino masses up to 185\GeV, setting stringent constraints on natural supersymmetry scenarios.}

\hypersetup{
pdfauthor={CMS Collaboration},
pdftitle={Search for electroweakinos in compressed-spectrum scenarios with low-momentum isolated tracks in proton-proton collisions at sqrt(s) = 13 TeV},
pdfsubject={CMS},
pdfkeywords={CMS, supersymmetry, higgsino dark matter, low-momentum track, multivariate classification}
}

\maketitle

Weak-scale $R$-parity conserving supersymmetry (SUSY)~\cite{Ramond:1971gb,Golfand:1971iw,Neveu:1971rx,Volkov:1972jx,Wess:1973kz,Wess:1974tw,Fayet:1974pd,Fayet:1976cr,Nilles:1983ge} is a well-motivated framework for physics beyond the standard model (SM). It provides a candidate for dark matter~\cite{Zwicky:1937zza,Rubin:1970zza}, resolves the large hierarchy problem~\cite{Barbieri:1987fn}, and enables the unification of gauge couplings. Models with 100\GeV-scale superpartners of the Higgs boson, known as higgsinos, are considered natural~\cite{Martin:1997ns} because they reduce the need for fine-tuning among SUSY parameters.

\begin{figure*}[!ht]
\centering
\includegraphics[width=0.3\textwidth]{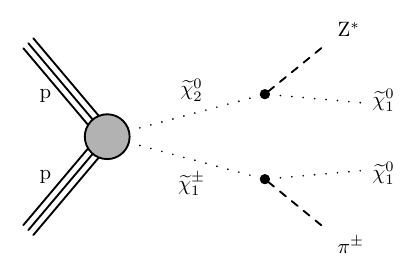}
\hfill
\includegraphics[width=0.3\textwidth]{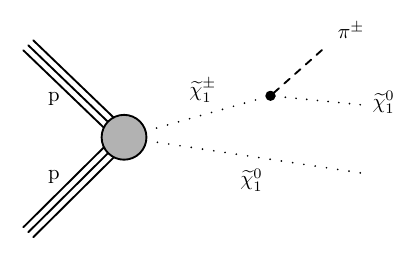}
\hfill
\includegraphics[width=0.3\textwidth]{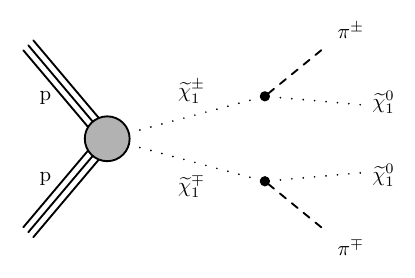}
\caption{Representative diagrams for the production of signal events. The signature considered in the analysis also involves one or more jets from initial state radiation with significant \pt in the final state, which are not shown in the figure. All processes are included in the interpretation of the search.}
\label{fig:signalfeynman}
\end{figure*}

Minimal SUSY models contain two Higgs doublets, the ratio of whose vacuum expectation values is given by $\tan\beta$. Electroweakino mass eigenstates consist of four neutralinos ($\PSGc^0_i$, $i=1$--4) and four charginos ($\PSGc^\pm_j$, $j=1,2$).  These arise from the mixing of the fermionic superpartners of the Higgs and electroweak gauge fields: the higgsinos, bino (superpartner of the \PB field), and winos (superpartners of the \PW fields), each associated with the mass parameters $\mu$, $M_1$, and $M_2$, which govern the electroweakino properties and mass spectrum. In the regime $\mu\ll M_1,M_2$, the lightest mass eigenstate \PSGczDo is a dominantly higgsino-like and stable neutralino, making it a viable dark matter candidate. It is accompanied by a nearly mass-degenerate second neutralino \PSGczDt and the lightest chargino \PSGcpmDo.
The mass splittings $\dmpm\equiv m(\PSGcpmDo)-m(\PSGczDo)$ and $\dmz\equiv m(\PSGczDt)-m(\PSGczDo)$ arise from electroweak radiative corrections and small mixing of the higgsinos with the wino and bino states, and typically range from 0.25 to 10\GeV~\cite{Nagata:2014wma}. The dominant production processes and decays are illustrated in Fig.~\ref{fig:signalfeynman}. For $\dmz\gtrsim1\GeV$, low-momentum (soft) leptons from electroweakino decays can be observed~\cite{ATLAS:2019lng,CMS:2021edw}. For $\dmpm\lesssim0.3\GeV$, the chargino lifetime is sufficiently long to yield a disappearing track signature~\cite{CMS:2020atg,ATLAS:2022rme,CMS:2023mny}. The intermediate range, 0.3--1\GeV, is particularly difficult to probe, as the chargino decays primarily to a single soft pion with a displacement of up to about 1\cm from the primary vertex (PV)~\cite{CMS-TDR-15-02}.

The analysis presented in this Letter is based on 138\fbinv of proton-proton collision data collected with the CMS detector at the CERN LHC during 2016--2018, targeting signal scenarios with $0.3<\dmpm<1\GeV$.  Sensitivity in this regime is achieved by exploiting the transverse momentum \pt and displacement of the pion track relative to the PV, as well as each event's missing transverse momentum vector \ptvecmiss and magnitude \ptmiss. The considered signal model corresponds to the large $\tan\beta$ regime~\cite{Nagata:2014wma}, in which the neutralino mass splitting satisfies $\dmz=2\dmpm$. In the scenario following Ref.~\cite{Mahbubani:2017gjh}, smaller \dmpm values yield longer \PSGcpmDo lifetimes and more pronounced displacement. Events are selected with \ptmiss exceeding 120\GeV at the trigger level and 300\GeV offline, along with at least one isolated track and no isolated electrons, muons, or photons. Rare decays of the \PSGczDt to soft electrons or muons are allowed, but the leptons have sufficiently small \pt to not invoke the lepton-based event vetoes. A parameterized neural network (NN) is trained to distinguish signal tracks from the dominant background. The background mainly consists of events with significant \ptmiss due to the presence of neutrinos, where soft tracks from sources unrelated to the hard process mimic the signal signature. A cut-based search targeting a similar phase space was previously reported by ATLAS~\cite{ATLAS:2024umc}, and recently a study with increased sensitivity based on an NN has been presented~\cite{ATLAS:2025lhc}.

The CMS apparatus~\cite{CMS:2008xjf,CMS:2023gfb} is a nearly hermetic detector, designed to trigger~\cite{CMS:2020cmk,CMS:2016ngn,CMS:2024aqx} on and identify electrons, muons, photons, and hadrons~\cite{CMS:2020uim,CMS:2018rym,CMS:2014pgm}. The  particle-flow (PF) algorithm~\cite{CMS:2017yfk} aims to reconstruct all detectable particles in an event, combining information provided by the silicon inner tracker and by the crystal electromagnetic and brass-scintillator hadron calorimeters, operating inside a 3.8\unit{T} superconducting solenoid, with data from the gas-ionization muon detectors embedded in the flux-return yoke outside the solenoid. The reconstructed particles are used to build \PGt leptons, jets, and the vector \ptvecmiss~\cite{CMS:2018jrd,CMS:2016lmd,CMS:2019ctu}. Jets are clustered using the anti-\kt algorithm~\cite{Cacciari:2008gp,Cacciari:2011ma} with a distance parameter of 0.4.

Monte Carlo (MC) samples of SM background processes, including top quark pair production, \wjets, and \zjets, are simulated at leading order (LO) in perturbative quantum chromodynamics using \MGvATNLO2.2.2~\cite{Alwall:2014hca,Alwall:2007fs}, with up to four additional partons in the matrix element. The \PYTHIA8.240~\cite{Sjostrand:2014zea} generator is used for parton showering, fragmentation, and hadronization, with matching between jets from the matrix element and those from the parton shower. The CUETP8M1~\cite{Khachatryan:2015pea} (CP5~\cite{CMS:2019csb}) underlying event tunes are used for background samples, along with NNPDF3.0 LO~\cite{Ball:2013hta} (NNPDF3.1 next-to-next-to-leading-order~\cite{Ball:2017nwa}) parton distribution functions for the 2016 (2017--2018) data-taking periods. Additional interactions in the same or adjacent bunch crossings (pileup) are generated with \PYTHIA and added to the hard-scatter event in all simulated samples, with the multiplicity distribution matching that observed in data. Signal events are simulated at LO using \PYTHIA8.240 and normalized to next-to-LO plus next-to-leading-logarithmic (NLO+NLL) precision.

Decays of the \PSGczDt to electrons or muons are set with a branching fraction of 5\% for each lepton flavor as a benchmark choice. The analysis sensitivity is only weakly affected by this assumption because no particle identification requirement is applied to the selected isolated track. For background events, the CMS detector response is modeled with the \GEANTfour~\cite{Agostinelli:2002hh}-based application \FullSim; to save computing resources, signal events are emulated with the CMS fast simulation engine \FastSim~\cite{Abdullin:2011zz,Giammanco:2014bza}.

Signal candidate tracks are reconstructed using the algorithm described in Ref.~\cite{CMS:2014pgm} and selected with $\pt<20\GeV$ and pseudorapidity $\abs{\eta}<2.4$. No requirement is imposed that the track be compatible with the PV, and no additional lower-\pt threshold is applied at the analysis level. The reconstruction retains an efficiency of about 10\% for \pt as low as 100\MeV, increasing to 80\% by 500\MeV and plateauing around 90\% by 1\GeV~\cite{CMS:2014pgm, Elmetenawee:2023xyl}. Criteria are applied to the track fit quality and the number of associated hits to ensure high purity of charged-particle reconstruction. Tracks must lie outside the reconstruction cone of all jets.  No dedicated pion identification is used. The separation of signal-like charged-particle tracks from PV-associated, pileup, and spurious tracks is performed by the classifier described below.  

Jets are required to have $\pt>30\GeV$ and $\abs{\eta}<5.0$, while hadronic \PGt-lepton candidates are identified from an alternate set of jets with $\pt>20\GeV$ and $\abs{\eta}<2.3$, using track, vertex, and calorimetric information and satisfying the tight working point (WP) of the \textsc{DeepTau} algorithm~\cite{CMS:2022prd}. Electrons (photons) are selected with $\abs{\eta}<2.5$ and $\pt>10$ (15)\GeV based on the medium WP~\cite{CMS:2020uim}. Muons are selected with $\pt>10\GeV$ and $\abs{\eta}<2.4$, with the loose identification WP~\cite{CMS:2018rym}. Electrons, photons, and muons are further required to satisfy a relative isolation requirement, in which the \pt sum of reconstructed particles within a cone of radius $\DR\equiv[(\Delta\eta)^2+(\Delta\phi)^2]^{1/2}=0.3$ around the object is less than 20\% of the object's \pt. The corresponding efficiencies are approximately 80\% for electrons and photons, and above 95\% for muons. Jets originating from the hadronization of \PQb quarks are identified using the \textsc{DeepJet} algorithm, with the medium WP corresponding to an efficiency in the 70--80\% range for \PQb jets, depending on jet kinematics~\cite{Sirunyan:2017ezt,ref:deepjet}.

The \ptvecmiss is computed as the negative vector \pt sum of all reconstructed PF candidates. A related quantity \mht, defined as the magnitude of the vector \pt sum of all jets with $\pt>30\GeV$ and $\abs{\eta}<5.0$, is also used to render the offline selection more consistent with the trigger requirement. Both measures are corrected to mitigate the effects of pileup and geometric dependencies of the jet energy response.

Events passing filters to reject anomalous \ptmiss arising from calorimeter noise, misreconstructed particles or vertices, or beam halo interactions are selected~\cite{CMS:2019ctu}. The baseline selection requires $\ptmiss>300\GeV$, $\mht>300\GeV$, and at least one jet with $\pt>100\GeV$, targeting signal events with significant initial state radiation. These requirements ensure the effective rejection of SM multijet events and a high trigger efficiency of 99\% or greater. Events must contain fewer than five jets, no \PQb-tagged jets, and a minimum $\Delta\phi$ of 0.4 radians between the vector \ptvecmiss and each of the four jets. Events containing photons, electrons, muons, or hadronically decaying \PGt leptons are rejected.

Background events remaining after these selection criteria include mainly \znunu, and \wlnu in which the lepton is out of acceptance or fails the identification criteria, yielding large \ptmiss from the neutrino(s). A large number of low-\pt tracks arise in each event from the underlying event, pileup interactions, and reconstruction failures, and are largely uncorrelated with the hard process. A rare population of tracks arises from the decay chain $\PW\to\PGt\PGn$.

\begin{figure*}[!ht]
\centering
\includegraphics[width=0.47\textwidth]{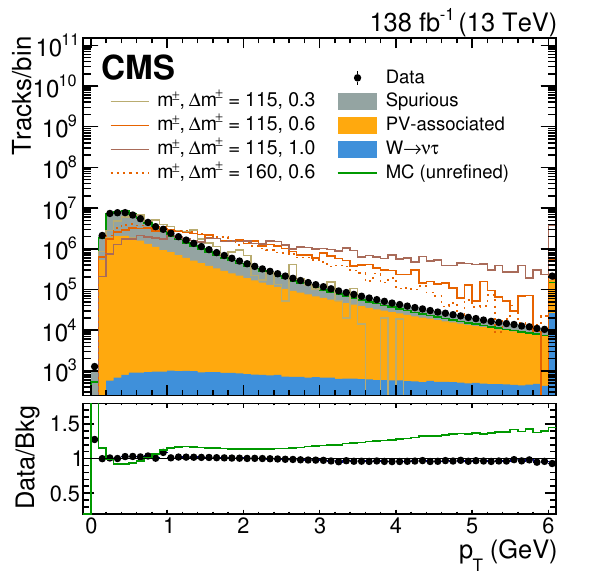}
\hfill
\includegraphics[width=0.47\textwidth]{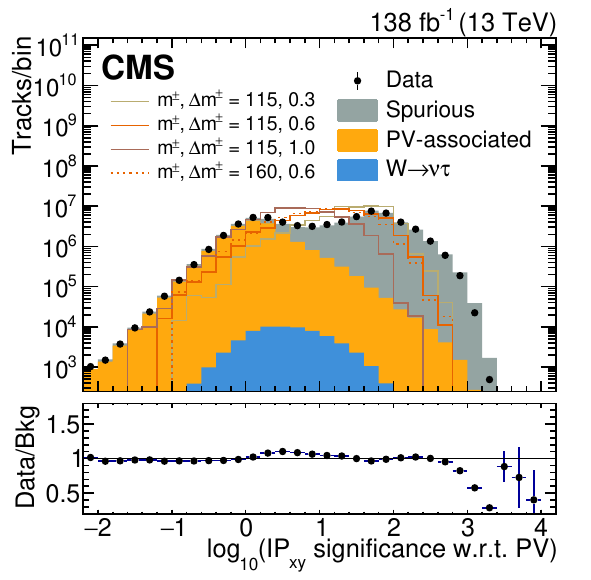}
\caption{Distributions of example kinematic track observables for data and simulation. The filled histograms show tracks in background events, with negligible statistical uncertainty. Simulated tracks matched to charginos from selected benchmark models, indicated by the chargino mass labels $m^{\pm}$ and \dmpm in \GeVns, are shown as colored open histograms normalized to the total background yield. The green line in the lower panel of the left plot indicates the ratio of the data to the background simulation before the refinement procedure is applied.}
\label{fig:signalcharacteristics}
\end{figure*}

Tracks arising from signal processes are typically isolated from jets and more closely aligned in direction with \ptvecmiss than  tracks in background events. The characteristic track \pt increases with larger \dmpm, while the transverse displacement, measured by the impact parameter, $\text{IP}_{xy}$, decreases, as shown by the colored open histograms in the main panels of Fig.~\ref{fig:signalcharacteristics}.
To suppress the overwhelming rate of background tracks, a deep parameterized NN is trained on approximately 4 million simulated reconstructed tracks, with training labels assigned using generator-level information. The NN has a three-layer architecture with 256 hidden units per layer, LeakyReLU activations~\cite{maas2013rectifier}, and preprocessing layers for input normalization. There are five output nodes corresponding to five track classes: signal tracks; secondary tracks from hadronic \PGt lepton decays in $\PW\to\tauh\PGn$ events; other secondary tracks from particles created at the PV; prompt tracks from the PV (together with the previous class referred to as ``PV-associated''); and spurious tracks. Spurious tracks lack generator-level matches and include contributions from pileup and misreconstructed tracks. The input features comprise 37 observables related to track kinematic properties, helix fit quality, impact parameter (IP) significance relative to each pileup vertex and the PV (where significance is defined as the measurement value divided by its uncertainty), and the event \ptmiss. The data-taking year is encoded as a parameter to account for varying detector conditions. The network is parameterized by \dmpm to capture correlations that evolve across the signal model phase space. Training is performed using the categorical cross-entropy loss. Track \pt provides comparatively larger discriminating power at large \dmpm, while displacement observables are more discriminating at low \dmpm due to longer \PSGcpm lifetime. Background tracks are assigned random values of \dmpm drawn from the inclusive signal distribution to avoid bias. The network is calibrated to produce per-track likelihood estimates. A logit transformation,
\begin{equation}
    \widetilde{P}\equiv\text{logit}\big(P\big)=\ln\left(\frac{P+\epsilon}{1-P+\epsilon}\right),
\end{equation}
is applied to the NN output score $P$ to enhance separation in the signal-like region for visualization and for use in the final statistical interpretation, where $\epsilon$ is a constant equal to $1\times10^{-6}$. 

Signal regions (SRs) are defined by evaluating the signal-class output node of the NN for tracks in each event at three values of input mass difference \dmpmin: 0.3, 0.6, and 1.0\GeV.  For each event, the track and \dmpmin with the highest score, denoted \pmaxsigsimp{\dmpmin}, are identified and the event is assigned to the corresponding \dmpmin category. This procedure defines three non-overlapping event categories, each further divided into bins of $\widetilde{P}_{\text{max}}$ as follows: (9.5--10, 10--11, 11--11.5, and $>$11.5) for $\dmpmin=0.3\GeV$; (8.75--9, 9--9.5, 9.5--10.5, and $>$10.5) for $\dmpmin=0.6\GeV$; and (7--7.5, 7.5--8, 8--8.75, and $>$8.75) for $\dmpmin=1.0\GeV$, for a total of 12 SRs.

Expected background yields are estimated using templates for each NN-defined background track class, derived from simulation and corrected using data. The modeling of the dominant PV-associated and spurious-track components is corrected and constrained with data in a cleaned Drell--Yan control region (DY CR), which is orthogonal to the SRs because of the requirement of two high-\pt muons. The cleaned DY CR is defined by selecting \zmumu events and removing the muons from the event record. Events are selected with two oppositely charged, isolated muons, each with $\pt>30\GeV$ and relative isolation $<$0.2. The dimuon invariant mass is required to be in the range 75--105\GeV and \pt exceeding 200\GeV. The \pt of the selected muons are vectorially added to \ptvecmiss, and the muons, as well as matching tracks, PF candidates, photons, and jets, are removed from the event record. The matching requires $\DR<0.4$ for jets and $\DR<0.05$ for all other objects.

The overall background rates of the PV-associated and spurious tracks in simulation are corrected using normalization factors derived in inclusive regions of the longitudinal IP, $d_z$, of candidate tracks in the cleaned DY CR. Tracks with $d_z<0.001\cm$ with respect to the PV are used to extract the scale factor for PV-associated tracks, while tracks with $d_z>0.63\cm$ define the region for spurious tracks. The resulting normalization factors are 0.92 for PV-associated tracks and 1.18 for spurious tracks. Simulated background samples are also corrected to improve agreement with the \ptmiss distribution observed in data. A weight is applied to simulated events, calculated by a linear fit to the data-to-simulation ratio as a function of \ptmiss in the range 300--1000\GeV, and events with $\ptmiss>1000\GeV$ are weighted using the value of the fit at $\ptmiss=1000\GeV$.

\begin{figure*}[!th]
\centering
\includegraphics[width=0.47\textwidth]{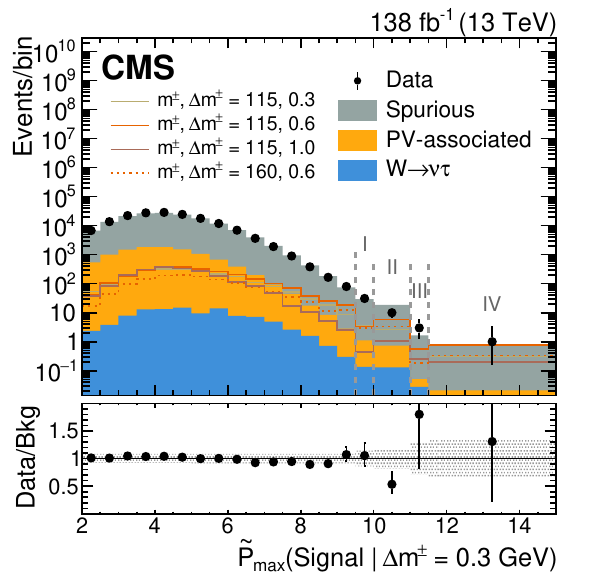}
\hfill
\includegraphics[width=0.47\textwidth]{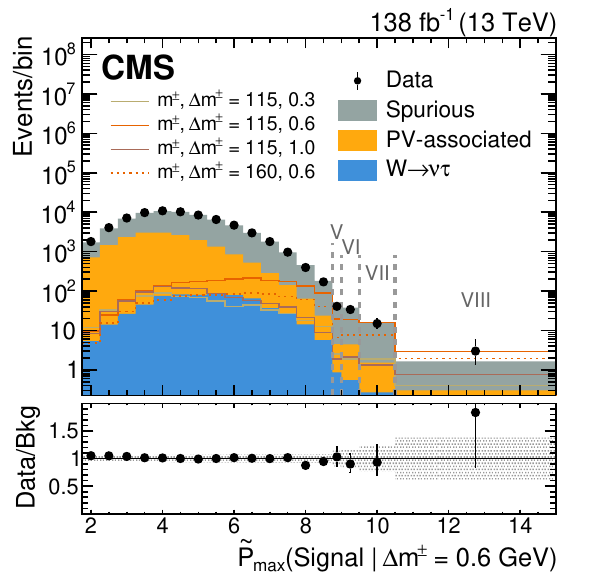} \\
\includegraphics[width=0.47\textwidth]{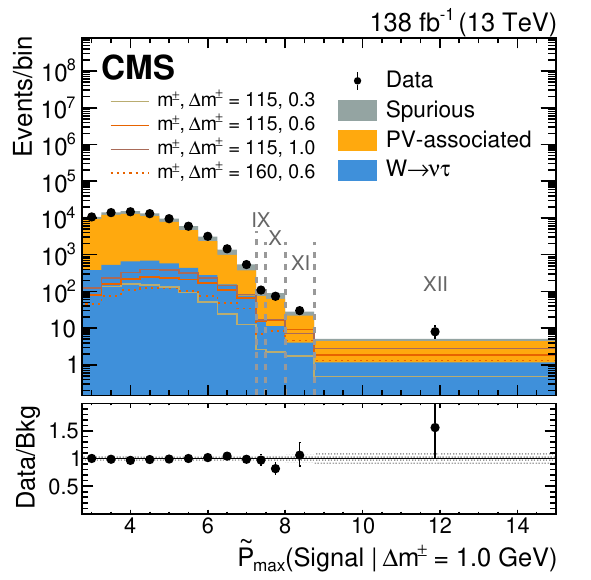}
\hfill
\includegraphics[width=0.47\textwidth]{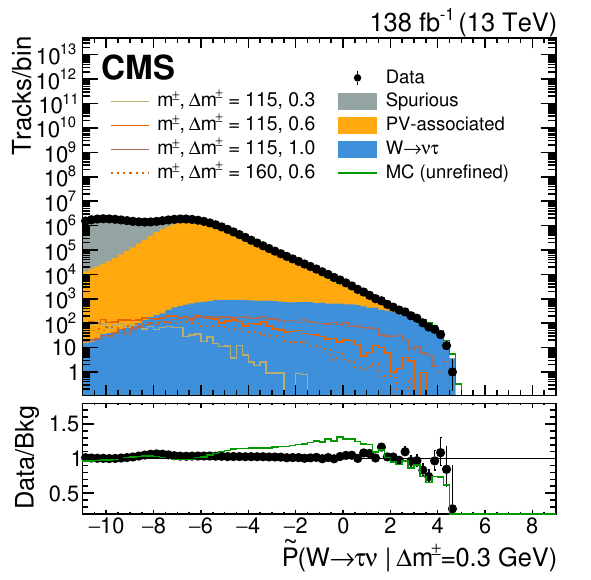}
\caption{The expected and observed yields as a function of the 0.3\GeV (upper left), 0.6\GeV (upper right), and 1.0\GeV (lower left) NN signal node, where the Roman numerals indicate the SR bin. In the upper panels, uncertainties shown are statistical only, while in the lower panels, shaded regions indicate the relative systematic uncertainty in the background yields and black vertical bars show the total statistical uncertainty in the ratio. The lower-right plot shows the distributions in the \PGt-lepton node of tracks after a selection of $\psig{0.3}>-2$, where the green line in the lower panel shows the ratio of the data to background simulation before the refinement procedure. The colored open histograms are as in Fig.~\ref{fig:signalcharacteristics}, except normalized to the integrated luminosity.
}
\label{fig:finalobservables}
\end{figure*}

Inaccuracies in the simulation of \pmaxsig{0.3,0.6,1.0}, arising from mismodeling of the track IPs and the \pt of soft tracks are corrected by a refinement procedure using a multivariate regression based on the Fast Perfekt methodology~\cite{Wolf:2024zkz}. An NN is trained and applied to morph simulated observables to better match those in data. For training, inclusive sets of tracks passing the baseline selection are taken from the cleaned DY CR in both data and simulation. The regression network outputs refined values of the observables by minimizing the maximum mean discrepancy loss~\cite{JMLR:v13:gretton12a} between the training data and the corrected simulation. The resulting corrections are applied to the track \pt and the uncertainties in the transverse and longitudinal IP, which serve as inputs to the classifier. The procedure improves the modeling of simulated tracks, particularly for PV-associated background and for \PGt-lepton-based signal proxy tracks, as shown in Figs.~\ref{fig:signalcharacteristics} (left) and~\ref{fig:finalobservables} (lower right). Differences between \FastSim and \FullSim in event- and track-level kinematics and displacement related variables are corrected using per-SR correction factors.

The uncertainty in the scale factors for the PV-associated and spurious backgrounds, accounting for potential mismodeling of the NN score shape in simulation, as well as potential bias stemming from the maximum-score criterion, is assessed to be 10\%. The uncertainty in the simulation shape, corrected through the refinement procedure, is evaluated by fitting polynomials to the ratio of data to the refined simulation in the signal-sensitive regions $\pmaxsig{0.3,0.6,1.0}>5$ of the cleaned DY CR and varies up to 10\%, depending on the SR. Additional uncertainties are applied to simulated events to account for potential mismodeling of the jet energy scale, pileup, trigger efficiency, and integrated luminosity~\cite{CMS-LUM-17-003,CMS-PAS-LUM-17-004,CMS-PAS-LUM-18-002,CMS-PAS-LUM-20-001}. An overall uncertainty of 10\% is assessed for the modeling of signal and \PGt-lepton events, based on the agreement between data and simulation in the \PGt-lepton-enriched CR, shown in Fig.~\ref{fig:finalobservables} in the lower-right panel. Uncertainties in the range 17--83\% in the signal yields are included to cover potential mismodeling in the \FastSim and statistical uncertainties in \FullSim-based signal samples used for calibrating the \FastSim.

\begin{figure*}[!ht]
\centering
\includegraphics[width=0.42\textwidth]{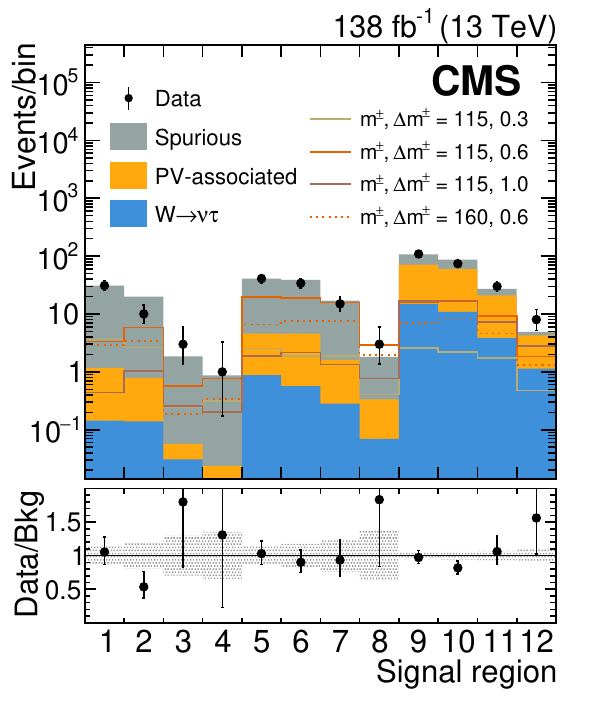}
\hfill
\includegraphics[width=0.52\textwidth]{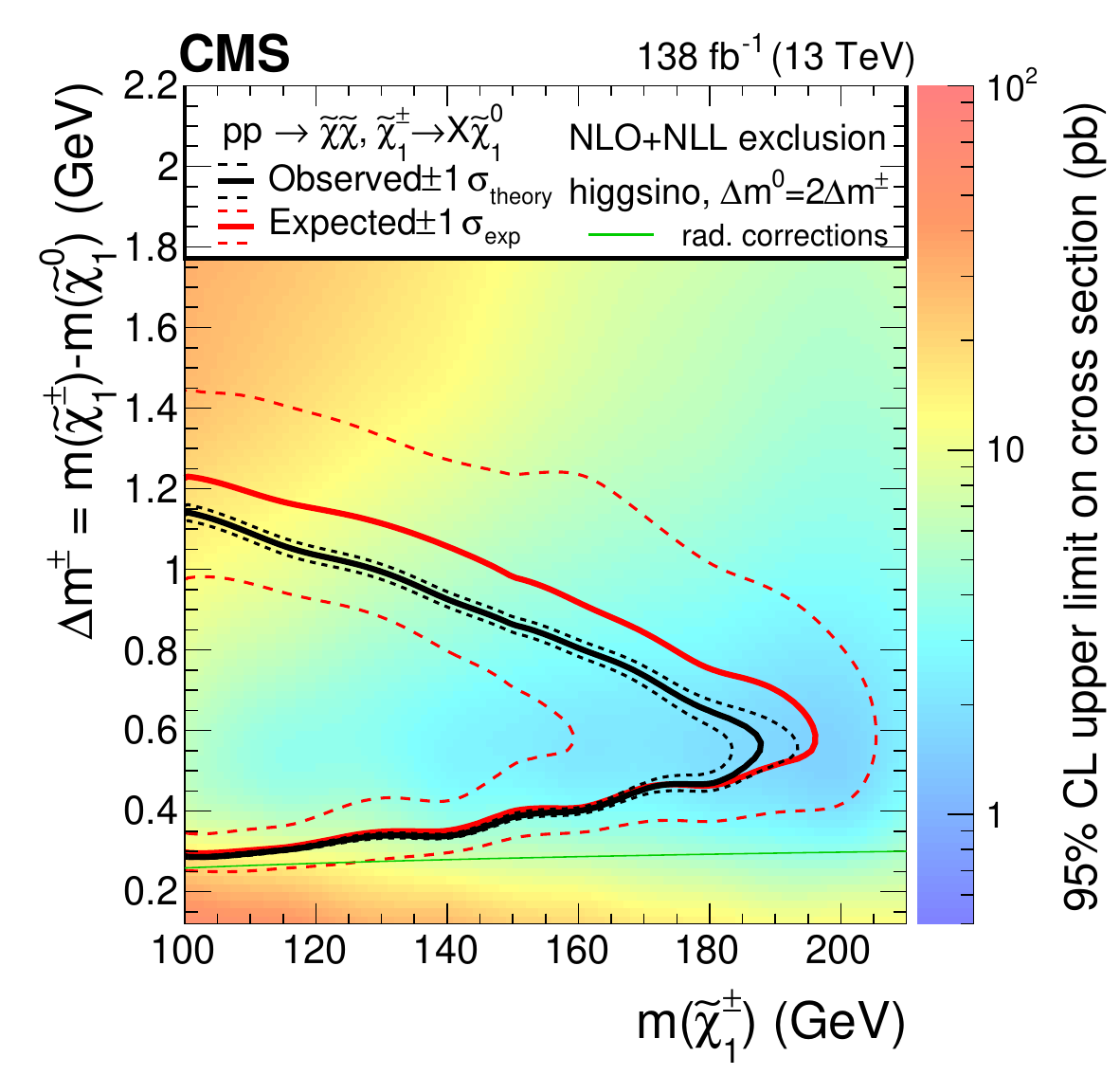}
\caption{The expected and observed data yields in the 12 SRs (left), where the uncertainties in the upper panel are statistical only, while in the lower panel the shaded region indicates the relative systematic uncertainty in the background yields and the black vertical bars show the total statistical uncertainty in the ratio. The colored open histograms are as in Fig.~\ref{fig:finalobservables}. The right plot shows the observed 95\% \CL upper limit on the cross section, accounting for all production modes of the higgsino model.  The expected (red) and observed (black) solid contours indicate the region excluded to the left of the curves, assuming NLO+NLL cross sections~\cite{Fuks:2012qx,Fuks:2013vua}. Dashed lines show the variation in the expected (observed) limits arising from experimental (theoretical) uncertainties. The chargino branching fractions and the minimum \dmpm allowed by radiative corrections (green line) are taken from Ref.~\cite{Ibe:2023dcu}.}
\label{fig:Limits}
\end{figure*}

The results are shown in Fig.~\ref{fig:finalobservables}, which displays the observed event counts compared to the stacked background expectations, along with superimposed yields of selected signal benchmark models. The SRs correspond to the four highest bins in each signal-class NN output distribution, which are shown in aggregate in Fig.~\ref{fig:Limits} (left). The results are consistent with the SM-only expectation.  The data are interpreted using a maximum likelihood fit. The likelihood is written
\begin{equation}
    \mathcal{L}(\mu_{s},\boldsymbol{\theta})=\prod_{i\in\mathrm{SR}}\mathrm{Pois}\big(n_i\,\big\vert\,\mu_{s}s_i(\boldsymbol{\theta})+b_i(\boldsymbol{\theta})\big)\prod_k\pi_k(\theta_k),
\end{equation}
where $n_i$ is the observed number of events in SR bin $i$, $s_i$ and $b_i$ are the expected signal and background yields, $\mu_{s}$ is the signal strength parameter, and $\boldsymbol{\theta}$ denotes the vector of nuisance parameters with index $k$ running over the sources of systematic uncertainty. The constraint terms $\pi_k(\theta_k)$ account for these uncertainties, with log-normal functions used for normalization uncertainties and Gaussian functions used for shape uncertainties. Correlations across bins are taken into account through the shared dependence of $s_i$ and $b_i$ on the nuisance parameters. The fit is performed simultaneously in the 12 SR bins. Upper limits at the 95\% confidence level (\CL) are derived on the inclusive electroweakino production cross section using a profile likelihood ratio test statistic based on the modified frequentist \CLs criterion with the asymptotic approximation~\cite{Junk:1999kv,Read:2002hq}.

The expected and observed limits are shown in Fig.~\ref{fig:Limits} (right) in the mass plane of the electroweakinos. The excluded region spans \dmpm values of 0.28--1.15\GeV for a \PSGcpmDo mass of 100\GeV and extends to 185\GeV for $\dmpm=0.55\GeV$.

In summary, a search for electroweakinos in compressed-spectrum scenarios has been performed using events containing a soft, isolated track and large missing transverse momentum. The results of the analysis are tabulated in Ref.~\cite{hepdata}. A parameterized neural network was employed to optimize sensitivity across a range of models not previously probed. No significant excess above the standard model prediction was observed. The resulting 95\% confidence level limits exclude chargino masses up to 185\GeV for a mass splitting between the chargino and lightest neutralino of 0.55\GeV, and mass splittings between  0.28 and 1.15\GeV for a chargino mass of 100\GeV, placing stringent constraints on natural supersymmetry scenarios.

\begin{acknowledgments}
We congratulate our colleagues in the CERN accelerator departments for the excellent performance of the LHC and thank the technical and administrative staffs at CERN and at other CMS institutes for their contributions to the success of the CMS effort. In addition, we gratefully acknowledge the computing centers and personnel of the Worldwide LHC Computing Grid and other centers for delivering so effectively the computing infrastructure essential to our analyses. Finally, we acknowledge the enduring support for the construction and operation of the LHC, the CMS detector, and the supporting computing infrastructure provided by the following funding agencies: SC (Armenia), BMBWF and FWF (Austria); FNRS and FWO (Belgium); CNPq, CAPES, FAPERJ, FAPERGS, and FAPESP (Brazil); MES and BNSF (Bulgaria); CERN; CAS, MoST, and NSFC (China); MINCIENCIAS (Colombia); MSES and CSF (Croatia); RIF (Cyprus); SENESCYT (Ecuador); ERC PRG and PSG, TARISTU24-TK10 and MoER TK202 (Estonia); Academy of Finland, MEC, and HIP (Finland); CEA and CNRS/IN2P3 (France); SRNSF (Georgia); BMFTR, DFG, and HGF (Germany); GSRI (Greece); MATE and NKFIH (Hungary); DAE and DST (India); IPM (Iran); SFI (Ireland); INFN (Italy); MSIT and NRF (Republic of Korea); MES (Latvia); LMTLT (Lithuania); MOE and UM (Malaysia); BUAP, CINVESTAV, CONACYT, LNS, SEP, and UASLP-FAI (Mexico); MOS (Montenegro); MBIE (New Zealand); PAEC (Pakistan); MSHE, NSC, and NAWA (Poland); FCT (Portugal); MESTD (Serbia); MICIU/AEI and PCTI (Spain); MOSTR (Sri Lanka); Swiss Funding Agencies (Switzerland); MST (Taipei); MHESI (Thailand); TUBITAK and TENMAK (T\"{u}rkiye); NASU (Ukraine); STFC (United Kingdom); DOE and NSF (USA).
\end{acknowledgments}

\bibliography{auto_generated}
\cleardoublepage \appendix\section{The CMS Collaboration \label{app:collab}}\begin{sloppypar}\hyphenpenalty=5000\widowpenalty=500\clubpenalty=5000\cmsinstitute{Yerevan Physics Institute, Yerevan, Armenia}
{\tolerance=6000
A.~Hayrapetyan, V.~Makarenko\cmsorcid{0000-0002-8406-8605}, A.~Tumasyan\cmsAuthorMark{1}\cmsorcid{0009-0000-0684-6742}
\par}
\cmsinstitute{Institut f\"{u}r Hochenergiephysik, Vienna, Austria}
{\tolerance=6000
W.~Adam\cmsorcid{0000-0001-9099-4341}, J.W.~Andrejkovic, L.~Benato\cmsorcid{0000-0001-5135-7489}, T.~Bergauer\cmsorcid{0000-0002-5786-0293}, M.~Dragicevic\cmsorcid{0000-0003-1967-6783}, C.~Giordano\cmsorcid{0000-0001-6317-2481}, P.S.~Hussain\cmsorcid{0000-0002-4825-5278}, M.~Jeitler\cmsAuthorMark{2}\cmsorcid{0000-0002-5141-9560}, N.~Krammer\cmsorcid{0000-0002-0548-0985}, A.~Li\cmsorcid{0000-0002-4547-116X}, D.~Liko\cmsorcid{0000-0002-3380-473X}, M.~Matthewman, I.~Mikulec\cmsorcid{0000-0003-0385-2746}, J.~Schieck\cmsAuthorMark{2}\cmsorcid{0000-0002-1058-8093}, R.~Sch\"{o}fbeck\cmsAuthorMark{2}\cmsorcid{0000-0002-2332-8784}, D.~Schwarz\cmsorcid{0000-0002-3821-7331}, M.~Shooshtari\cmsorcid{0009-0004-8882-4887}, M.~Sonawane\cmsorcid{0000-0003-0510-7010}, W.~Waltenberger\cmsorcid{0000-0002-6215-7228}, C.-E.~Wulz\cmsAuthorMark{2}\cmsorcid{0000-0001-9226-5812}
% ...
\par}
\vskip\cmsinstskip
\dag:~Deceased\\
$^{1}$Also at Yerevan State University, Yerevan, Armenia\\
$^{2}$Also at TU Wien, Vienna, Austria\\
% ...
\end{sloppypar}
\end{document}